\documentclass[epj,nopacs]{svjour}

\usepackage{amsmath,amssymb,feynmp
}
\usepackage[dvips]{graphicx}

\usepackage[a4paper,dvips,draft=false]{hyperref}

\newcommand{\nodata}{\multicolumn{1}{|c|}{---}}
\newcommand{\eminbr}[1]{\widehat{E}_{\rm{min}}^{#1}}
\newcommand{\meminbr}[1]{\widehat{E}_{\rm{rem}}^{#1}}
\begin{document}
\begin{fmffile}{feyndiag}

\title{Z boson decay to photon plus Kaluza-Klein %
 graviton: large extra dimensional bounds}

\author{Benjamin C.~Allanach \and Jordan P.~Skittrall}

\institute{Department of Applied Mathematics and Theoretical Physics,
  Centre for Mathematical Sciences, University of Cambridge,
  Wilberforce Road, Cambridge, CB3 0WA, United Kingdom
  \email{B.C.Allanach@damtp.cam.ac.uk, J.P.Skittrall@damtp.cam.ac.uk}}

\mail{Jordan P.~Skittrall}

\abstract{%
We consider the phenomenology of the decay of a Z~boson into a photon
and a Kaluza-Klein 
excitation of the graviton in the ADD model. Using LEP data, we obtain
an upper bound on the branching ratio corresponding to this process of
$\sim 10^{-11}$.
We also
investigate energy profiles of the
process.
}


\titlerunning{$Z\to\gamma\mathcal{G}$: large extra dimensional bounds}
\authorrunning{Benjamin C.\ Allanach and Jordan P.\ Skittrall}

\maketitle

\section{Introduction}

%
Nieves and Pal~\cite{NiP2005} have constructed a theoretical argument to
predict the decay width of a Z~boson to a photon and a graviton.
The ADD model \cite{ArDD1998,AnADD1998,Ko2002,CrIM2002} predicts a ``tower'' of massive Kaluza-Klein
excitations of the graviton (both massive spin-2 gravitons and massive
spin-0 gravi-scalars) when the model is viewed from a
four-dimensional perspective \cite{GiRW1998,HaLZ1998}. In a previous paper with K.~Sridhar
\cite{AlSS2007}, we extended the theoretical argument of Nieves and
Pal to predict the decay width of a Z~boson to a photon
and a Kaluza-Klein graviton state in an ADD model where the extra
dimensions are toroidally compactified with a common compactification
radius.
\setlength{\unitlength}{1cm}
\begin{figure}
\begin{center}
\begin{fmfgraph*}(6,4)
\fmfleft{l1}
\fmfright{r1,r2}
\fmf{zigzag,label=$Z$,l.side=left}{l1,i1}
\fmf{photon,label=$\gamma$}{i1,r2}
\fmf{dbl_wiggly,label=$\mathcal{G}$}{i1,r1}
\fmfblob{1cm}{i1}
\end{fmfgraph*}
\caption{The decay of a Z~boson to a photon and a Kaluza-Klein
  graviton is one-loop at leading order.}\label{fig:genericfeyndiagram}
\end{center}
\end{figure}
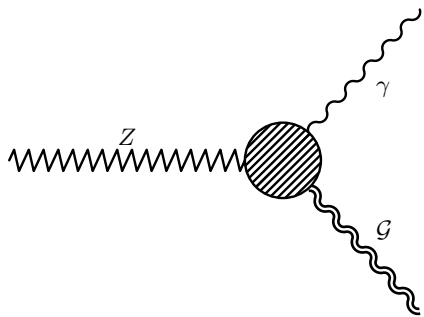
We now extend that work by using data from the L3 experiment at LEP1
\cite{Acetal1997} to estimate
the bounds on the size of such extra dimensions that can be achieved
by considering this process. The bounds we obtain are comparable with,
but not stronger than, those obtained by considering the tree-level
processes $e^+ e^- \to \gamma \mathcal{G}$ \cite{ADLOLEWG2004} and
$p\bar{p}\to \mathcal{G} +{\rm jet}$ \cite{Abetal2006,Abetal2003}. (In
this paper, we use
$\mathcal{G}$ to denote any Kaluza-Klein excitation of the graviton --
massive graviton or gravi-scalar -- in the tree-level processes
the gravi-scalar contribution is negligible, but in the one-loop
Z decay process it dominates.)
By using the bounds on the size of such extra dimensions obtained from
the processes $e^+ e^- \to \gamma \mathcal{G}$ and $p\bar{p}\to
\mathcal{G} +{\rm jet}$, and, for the case of two
extra dimensions, from inverse square law experiments \cite{HoKHAGSS2004,Ad2002}, it is
therefore possible to predict stronger bounds than those measured by
experiment on the contribution from
the process $Z\to\gamma\mathcal{G}$ to the decay $Z \to \gamma +
\textrm{missing }E_t$. The weakest such bound constrains the
branching ratio to around the $10^{-11}$ level.

We also investigate energy spectra of events from this decay
channel. These spectra may 
prove useful in model
discrimination, possibly indicating an experiment to be performed were
ADD discovered. It is particularly notable that the shape of the
spectra for $n=2$ extra dimensions differs from those for $n>2$ extra
dimensions.

\section{Existing bounds and bounds from LEP1 data on Z~decay}

Existing particle physics bounds on the size of extra dimensions come
from consideration of the tree-level processes $e^+ e^- \to \gamma
\mathcal{G}$ \cite{ADLOLEWG2004} and $p\bar{p}\to \mathcal{G}+{\rm
  jet}$ \cite{Abetal2006,Abetal2003}. In our paper with K.~Sridhar
\cite{AlSS2007}, we showed that
\begin{small}
\begin{multline}
\Gamma_{\rm{tot}} = \frac{\alpha^2 G M_Z^{3+n}R^n
    \pi^{n/2}}{72\pi^2} \ \times
\\
 \shoveleft{\times \left[
  0.00088\left( \frac{7 \cdot 5!}{\Gamma(\frac{n}{2}+6)} + \frac{3
    \cdot \frac{n}{2} \cdot 5!}{\Gamma(\frac{n}{2}+7)} \right) +
  \phantom{ \left\{ \sum_{j=0}^{\infty}\right\}} \right.}
\\ 
\shoveleft{ +  0.27 \left\{ \sum_{j=0}^{\infty}
    \frac{1}{(j+2)(j+3)(j+4)}\right.}
\\
\shoveright{\left.\left( \frac{7
    \cdot (j+5)!}{\Gamma(\frac{n}{2}+j+6)}+\frac{3 \cdot \frac{n}{2}
    \cdot (j+5)!}{\Gamma(\frac{n}{2} + j + 7)} \right) \right\} +}
\\
 \shoveleft{+ 21 
\left\{ \sum_{i=0}^{\infty} \sum_{j=0}^{\infty}
\frac{1}{(i+2)(i+3)(i+4)(j+2)(j+3)(j+4)} \times \right.} 
\\
\shoveright{ \left.
\times  \left(\frac{7
    \cdot (i+j+5)!}{\Gamma(\frac{n}{2}+i+j+6)}+\frac{3 \cdot \frac{n}{2}
    \cdot (i+j+5)!}{\Gamma(\frac{n}{2} + i+j + 7)}
  \right)\right\} +} 
\\
\shoveleft{+ \frac{3}{2} \frac{(n-1)}{(n+2)}
 \left\{
  \frac{330}{\Gamma(\frac{n}{2}+4)} - \frac{63 \cdot
  \frac{n}{2}}{\Gamma(\frac{n}{2}+5)} + \frac{13 \cdot \frac{n}{2}
  \cdot (\frac{n}{2}+1 )}{\Gamma(\frac{n}{2}+6)}\right. -} 
\\
\shoveright{
- \frac{0.97 \cdot
  \frac{n}{2} \cdot (\frac{n}{2}+1) \cdot
  (\frac{n}{2}+2)}{\Gamma(\frac{n}{2}+7)} -} 
\\
\left.  \phantom{\sum_{j=0}^{\infty}}
 \left. - \frac{0.078 \cdot
  \frac{n}{2} \cdot (\frac{n}{2}+1) \cdot
  (\frac{n}{2}+2) \cdot (\frac{n}{2}+3)}{\Gamma(\frac{n}{2}+8)} \right\}  \right] ,
\label{eq:fullwidth}
\end{multline}
\end{small}
where $\Gamma_{\rm{tot}}$ is the total decay width of the Z~boson to a
photon and any single Kaluza-Klein excitation of the graviton, $n$ is
the number of extra dimensions, all of which are toroidally compactified with
common radius $R$, $G$ is Newton's constant in four dimensions,
$\alpha$ is the fine-structure constant and $M_Z$ is the mass of the
Z~boson. The width has contributions from decay to a photon and a
spin-0 gravi-scalar (the section with the $(3/2)(n-1)/(n+2)$
coefficient) and from decay to a photon and a spin-2 Kaluza-Klein
graviton excitation (the remainder).

Instead of being expressed in terms of a compactification radius $R$,
it is equivalently possible to express the decay width (and
compactification scale) in terms of a mass $M_D$, which is defined
such that
\begin{equation}
8\pi R^n M_D^{n+2} G = 1 \, ,
\end{equation}
where again $G$ is Newton's constant in four dimensions. We present
bounds using both measures of compactification scale.

It is possible to obtain experimental bounds on the
width of the decay $Z\to \gamma + \textrm{missing }E_t$ either by direct event
selection, or by subtracting from the total Z~width the sum of the
widths of the ``visible'' Z decays. The method of direct event
selection leads to much
stronger bounds with current data, and so it is the method we use in
this paper. With both methods (and in the case of the process  $e^+ e^- \to
\gamma \mathcal{G}$), there is a Standard Model background to
experimental data from the
process $e^+e^-\to \gamma \nu \bar{\nu}$
\cite{MaO1978,GaGR1979,BaRS1981,JaKW1998}. The experimental analyses
take this background at tree-level into account to produce 95\%
confidence bounds,
which we use. Of the one-loop corrections to the background, those
involving the
process $Z\to \gamma \nu \bar{\nu}$ are of particular relevance to
the Z~decay case, but the branching ratio for the process $Z\to \gamma
\nu \bar{\nu}$ \cite{HePTT1999} is (at $7.16 \times 10^{-10}$) about four orders of magnitude
smaller than the L3 experimental bound on the process
$Z\to \gamma + \textrm{missing }E_t$ \cite{Acetal1997},
and so the process $Z\to \gamma \nu \bar{\nu}$ is negligible when analysing current experimental bounds on the process
$Z\to \gamma + \textrm{missing }E_t$.

In order to consider experimental data to derive a bound on the
magnitude of the radius of the extra dimensions $R$, we need to take
into account that there will be a cut specifying the minimum photon
energy, $E_{\rm min}$. We can put this into our summation/integration over the
Kaluza-Klein states when deriving the overall width, to get
\begin{small}
\begin{multline}
\Gamma_{\rm{tot}} =  \frac{\alpha^2 G M_Z^{3+n}R^n
  \pi^{n/2}}{72\pi^2} \ \times
\\
  \shoveleft{\times \left[ \left\{ 0.00088 \times \phantom{\frac{1}{j}}
  \right.\right.}\\
  \shoveleft{\times \left[ 10
  \sum_{p=0}^{5}\left( \frac{5!}{(5-p)! \, \Gamma(\frac{n}{2}+p+1)}
  \eminbr{5-p} \meminbr{\frac{n}{2}+p} \right) -
  \right.} \\
\shoveright{\left. \phantom{=\Bigg[ \Bigg\{ \times \Bigg[} - 3\sum_{p=0}^{6}\left(
  \frac{6!}{(6-p)! \,
  \Gamma(\frac{n}{2}+p+1)} \eminbr{6-p} \meminbr{\frac{n}{2}+p}
  \right) \right] +} \\
\shoveleft{ +0.27 \sum_{j=0}^{\infty}
  \frac{1}{(j+2)(j+3)(j+4)} \times} \\
 \times \left[ 10
  \sum_{p=0}^{5+j} \left( \frac{(5+j)!}{(5+j-p)! \,
  \Gamma(\frac{n}{2}+p+1)} \eminbr{5+j-p} \meminbr{\frac{n}{2}+p}
  \right) - \right. \\
 \left. 
\shoveright{- 3
  \sum_{p=0}^{6+j} \left( \frac{(6+j)!}{(6+j-p)! \,
  \Gamma(\frac{n}{2}+p+1)} \eminbr{6+j-p} \meminbr{\frac{n}{2}+p}
  \right) \right] +} \displaybreak[2]\\
 \shoveleft{ +21\sum_{i=0}^{\infty}\sum_{j=0}^{\infty}\frac{1}{(i+2)(i+3)(i+4)(j+2)(j+3)(j+4)}
  \times }\\
\times
  \left[ 10\sum_{p=0}^{5+i+j} \left( \frac{(5+i+j)!}{(5+i+j-p)! \,
  \Gamma(\frac{n}{2}+p+1)} \eminbr{5+i+j-p} \meminbr{\frac{n}{2}+p}
  \right) - \right. \\
 \shoveright{\left. \left. - 3
  \sum_{p=0}^{6+i+j} \left( \frac{(6+i+j)!}{(6+i+j-p)! \,
  \Gamma(\frac{n}{2}+p+1)} \eminbr{6+i+j-p} \meminbr{\frac{n}{2}+p}
  \right) \right] \right\} +} \displaybreak[2]\\
 \shoveleft{ + \frac{1}{4}\frac{n-1}{(n+2)\Gamma(\frac{n}{2})} \times} \\
  \shoveleft{\times \left\{
  \frac{330}{\frac{n}{2}}
  \meminbr{\frac{n}{2}}-\frac{910}{\frac{n}{2}+1}\meminbr{\frac{n}{2}+1}
   +\frac{800}{\frac{n}{2}+2} \meminbr{\frac{n}{2}+2}  -
  \right.} \\
-  \frac{180}{\frac{n}{2}+3}\meminbr{\frac{n}{2}+3} -
  \frac{26}{\frac{n}{2}+4} \meminbr{\frac{n}{2}+4} -
  \frac{10}{\frac{n}{2}+5} \meminbr{\frac{n}{2}+5} -
  \\  \left. \left. -
\frac{0.74}{\frac{n}{2}+6} \meminbr{\frac{n}{2}+6} +
  \frac{0.078}{\frac{n}{2}+7} \meminbr{\frac{n}{2}+7}
  \right\}\right],\label{eq:energydependentwidth}
\end{multline}
\end{small}
where we have defined $\widehat{E}_{\rm{min}} \equiv
2E_{\rm{min}}/M_Z$ and $\widehat{E}_{\rm{rem}} \equiv 1 -
\widehat{E}_{\rm{min}}$ for notational brevity.
(This reduces to equation~\eqref{eq:fullwidth} in the case $E_{\rm{min}}=0$.)

We can apply this formula to the data of reference~\cite{Acetal1997},
to obtain the bounds on the size of the extra dimensions given in the
second and third columns of
table~\ref{tab:bounds}\footnote{Reference~\cite{Acetal1997} plots
  upper limits on the branching ratio for a range of values of
  $E_{\rm{min}}$. The bounds on the size of the extra dimensions we
  obtain are the strongest that arise
  from applying the formula of
  equation~\eqref{eq:energydependentwidth} across the range of values
  of $E_{\rm{min}}$. For each number of extra dimensions $n$, the
  value of $E_{\rm{min}}$ that leads to the strongest bound is 30.8~GeV.}.

\begin{table}
\begin{center}
\begin{footnotesize}


\begin{tabular}{|c|l|l|l|l|}
\hline
 & \multicolumn{1}{p{1cm}}{L3} &
 \multicolumn{1}{|p{1cm}|}{LEP} &
 \multicolumn{1}{|p{1cm}|}{CDF} & \multicolumn{1}{|p{1cm}|}{ISL} \\
\cline{2-5}
$n$ & \multicolumn{1}{|p{1.3cm}|}{$M_D$ (TeV) $>$} & \multicolumn{1}{|p{1.3cm}|}{$M_D$ (TeV) $>$} & \multicolumn{1}{|p{1.3cm}|}{$M_D$ (TeV) $>$} &
 \multicolumn{1}{|p{1.3cm}|}{$M_D$ (TeV) $>$} \\\hline
2 & $0.18$ & $1.6$ & $1.18$ & {\boldmath
 $1.9$} \\
3 & $0.16$ & {\boldmath $1.2$} & $0.99$ & \nodata \\
4 & $0.14$ & {\boldmath $0.94$} & $0.91$ & \nodata \\
5 & $0.13$ & $0.77$ & {\boldmath $0.86$} & \nodata \\
6 & $0.12$ & $0.66$ & {\boldmath $0.83$} & \nodata \\
\hline
\end{tabular}

\vspace{0.3em}

\begin{tabular}{|c|p{1.5cm}|p{1.6cm}|p{1.7cm}|l|}
\hline
 & \multicolumn{1}{p{1cm}}{L3} &
 \multicolumn{1}{|p{1cm}|}{LEP} &
 \multicolumn{1}{|p{1cm}|}{CDF} & \multicolumn{1}{|p{1cm}|}{ISL} \\
\cline{2-5}
$n$ & $R$ (mm) $<$ & $R$ (mm) $<$
 & $R$ (mm) $<$ & \multicolumn{1}{|p{1.5cm}|}{$R$ (mm) $<$} \\\hline
2 & $15$ & $0.19$ & $0.35$ & {\boldmath  $0.13$} \\
3 & $7.4 \times 10^{-5}$ & {\boldmath $2.6 \times 10^{-6}$}
 & $3.6 \times 10^{-6}$ & \nodata \\
4 & $1.8 \times 10^{-7}$ & {\boldmath
  $1.1 \times 10^{-8}$}
 & $1.1 \times 10^{-8}$ & \nodata \\
5 & $5.0 \times 10^{-9}$ & $4.1 \times
 10^{-10}$ & {\boldmath $3.5 \times 10^{-10}$} & \nodata \\
6 & $4.6 \times 10^{-10}$ & $4.6 \times
 10^{-11}$ & {\boldmath $3.4 \times 10^{-11}$} & \nodata \\
\hline
\end{tabular}

\end{footnotesize}

\caption{Bounds on the scales of the extra dimensions (lower limits on
  the gravity scale $M_D$ and upper limits on the radius $R$), for $n=2$ to
  $n=6$ extra dimensions, for (L3) L3 Z decay data \cite{Acetal1997} (LEP)
Combined LEP $e^+ e^-
 \to \gamma \mathcal{G}$ data \cite{ADLOLEWG2004}  (CDF) CDF Run II $p\bar{p}\to\mathcal{G}+{\rm jet}$
 data \cite{Abetal2006} (ISL) Inverse square law experiment data \cite{HoKHAGSS2004}. All limits correspond to a $95\%$ confidence
  level. The strongest bound for each value of $n$ is shown in bold.}\label{tab:bounds}
\end{center}
\end{table}

In addition to particle physics bounds, a stronger experimental bound
can be obtained for $n=2$ extra dimensions from inverse square law
experiments~\cite{HoKHAGSS2004}. (Bounds cannot be directly taken from
reference \cite{HoKHAGSS2004} for $n>2$ extra dimensions as the bounds derived require the
extra dimensions to be asymmetrically sized.) There are also
estimates of astrophysical bounds
using the temperature profile of the observed collapse of SN1987A (an
upper bound is set on the amount of graviton emission as this would
affect the resultant temperature)
\cite{ArDD1998a,CuP1999,BaHKZ1999,HaPRS2000}, which depending on the
assumptions made give estimates of $M_D \gtrsim 30$~TeV to $M_D
\gtrsim 130$~TeV for
$n=2$, and $M_D \gtrsim 2.0$~TeV to $M_D \gtrsim 9.3$~TeV for $n=3$
(the bounds are comparable with or weaker than
the experimental bounds for $n>3$). There are also cosmological
arguments that lead to estimated bounds on the size of extra
dimensions \cite{BaD1998,HaS1999}.

We see that the bounds on the scale of the extra dimensions derived
from the Z~decay process are weaker than other bounds. It is possible
therefore for us to use the stronger bounds on the scale to estimate
upper bounds
on the branching ratio of the Z~boson to a photon and a Kaluza-Klein
graviton/gravi-scalar, using equation~\eqref{eq:fullwidth} and reference~\cite{Yaetal2006}.
\begin{table}
\begin{center}
\begin{tabular}{|c|l|}
\hline
$n$ & Branching ratio bound \\
\hline
2 & $1 \times 10^{-11}$ \\
3 & $6 \times 10^{-12}$ \\
4 & $2 \times 10^{-12}$ \\
5 & $3 \times 10^{-13}$ \\
6 & $3 \times 10^{-14}$ \\
\hline
\end{tabular}
\caption{Limits on the total branching ratio of the
  Z~boson to a photon and a Kaluza-Klein graviton/gravi-scalar at the
  strongest of each of the $95\%$ confidence bounds on the mass scale
  $M_D$ from table~\ref{tab:bounds} (i.e.~the bounds given in bold).}\label{tab:brfrombounds}
\end{center}
\end{table}
\begin{figure}
\begin{center}
\includegraphics[width=1\columnwidth]{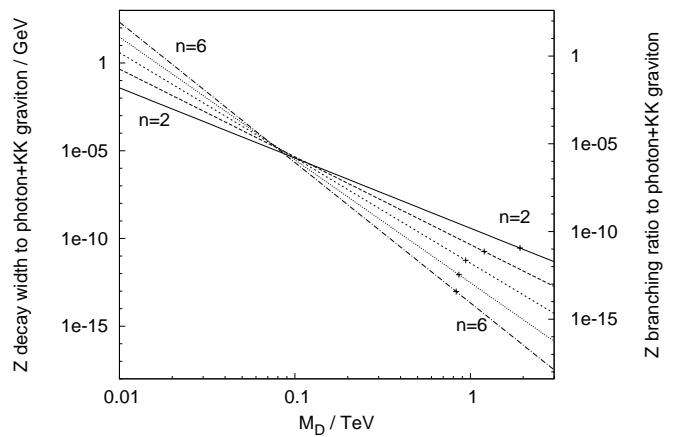}
\caption{$\Gamma(Z\to\gamma\mathcal{G})$ and
  ${\rm BR}(Z\to\gamma\mathcal{G})$ as functions of of the mass
  scale $M_D$ of the extra dimensions, for $n=2$ to $n=6$ extra
  dimensions (consecutive values of $n$ correspond to adjacent lines
  in the plot). The crosses plotted show the bounds on the decay width
  and branching ratio obtained by considering the strongest
  experimental $95\%$ confidence
  bounds on $M_D$ (i.e.~the bold bounds from
  table~\ref{tab:bounds}).}\label{fig:analyticcalculatebw}
\end{center}
\end{figure}
Table~\ref{tab:brfrombounds} gives such an estimate. The estimate is
also plotted in figure~\ref{fig:analyticcalculatebw}, which shows how
the decay widths and branching ratios for the process depend upon the
scale of the extra dimensions (so that the figure is, in effect, a
plot of equation~\eqref{eq:fullwidth}). We note that the largest
branching ratio (in the $n=2$ case) is of the order $10^{-11}$, so
that we should not expect to see any such events in a Giga-Z collider,
if only $10^9$ events were collected, indicating a need for an
experiment of higher luminosity.

\section{Energy profiles of the process}

Extra dimensional models are often distinguishable from other ``new
physics'' models by noting that the multiple Kaluza-Klein states
predicted by the extra dimensional model give a different, usually
softer, energy distribution from those corresponding to models
predicting single new states. With this in mind, we investigate the
energy distribution of the process $Z\to\gamma\mathcal{G}$. It is
necessary to recall the caveat that the low branching ratio means a
very high Z luminosity will be
required to see such a distribution. The likely main use of such profiles
would therefore be in model discrimination subsequent to a signal seen
elsewhere. An analysis of experimental data would also need to take
into account the energy profile of the Standard Model background
\cite{BaRS1981,JaKW1998}, in addition to any other event selection
considerations.

The differential decay width $d\Gamma/dE$, where $E$ is the energy of
the photon in the centre-of-mass frame, is equal to
$-d\Gamma(E_{\rm{min}})/dE_{\rm{min}}$, where $\Gamma(E_{\min})$ is as
given in equation~\eqref{eq:energydependentwidth} (the minus sign
comes from the minimum energy cut corresponding to a lower bound in
the integration over the Kaluza-Klein mass states).

\begin{figure}
\begin{center}
\includegraphics[width=0.9\columnwidth]{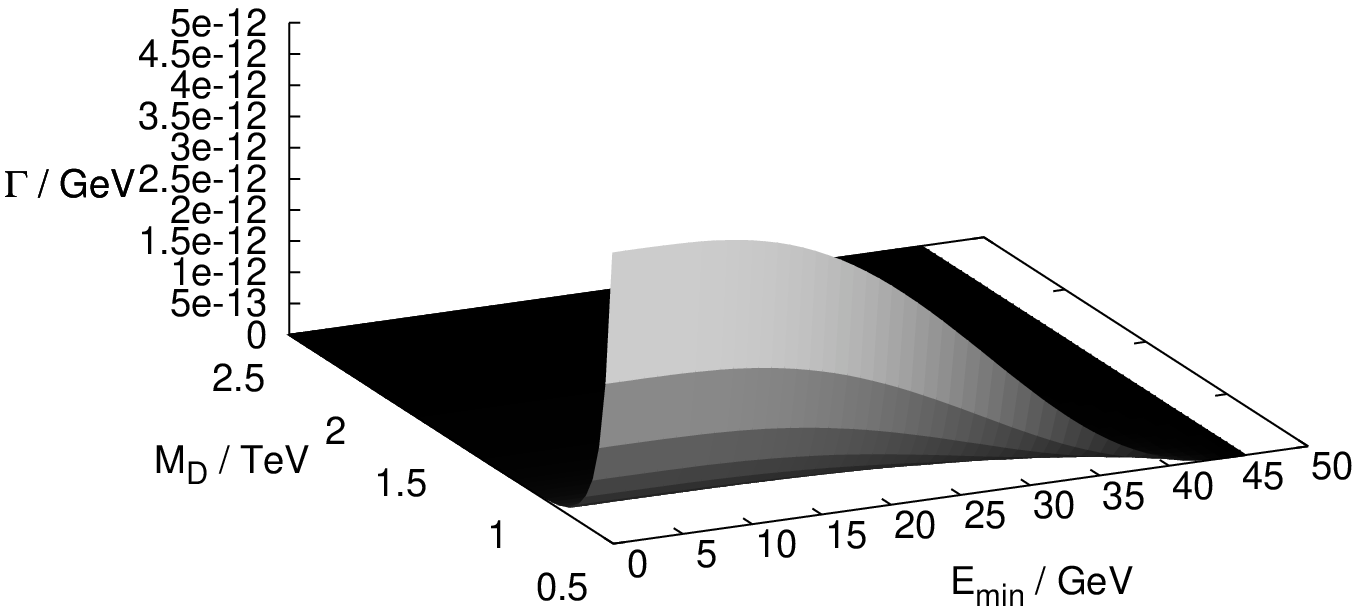}
\\
(e) $n=6$
\caption{Decay widths for the process $Z\to\gamma\mathcal{G}$, for
  mass scales $M_D=0.5$~TeV to $M_D=2.5$~TeV of the extra dimensions,
  with a photon energy cut $E_{\rm{min}}$, in $n=2$ to $n=6$ extra dimensions.}\label{fig:gammaemdgraphs}
\end{center}
\end{figure}

\begin{figure}
\begin{center}
\includegraphics[width=0.9\columnwidth]{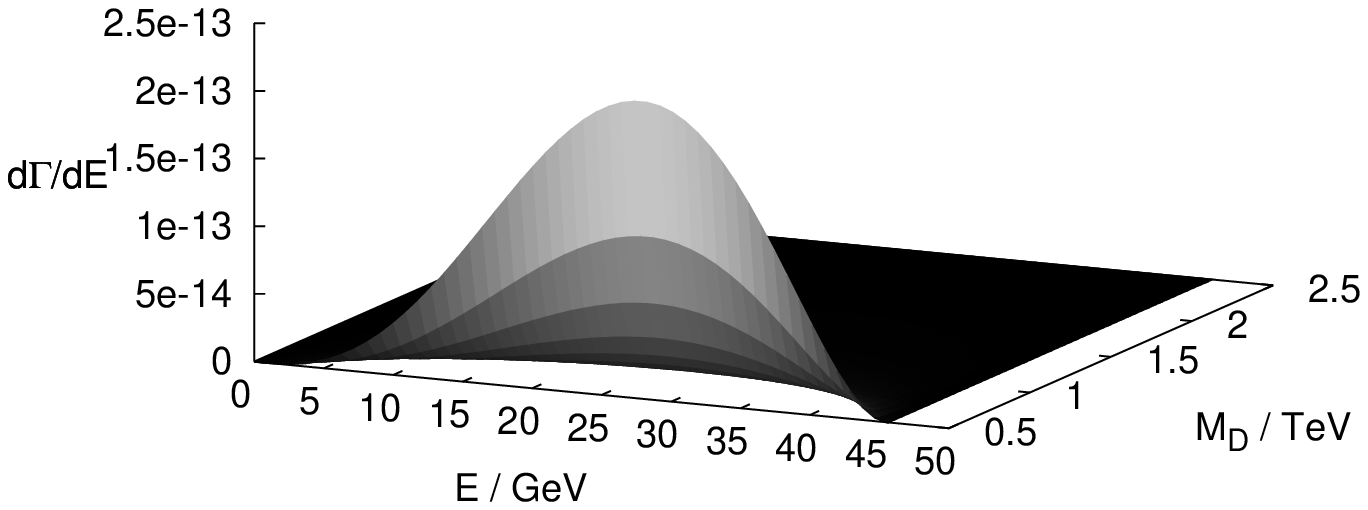}
\\
(e) $n=6$
\caption{Differential decay widths $d\Gamma / dE$ for the process
  $Z\to\gamma\mathcal{G}$, where $E$ is the energy in the
  centre-of-mass frame of the photon
  produced, for
  mass scales $M_D=0.5$~TeV to $M_D=2.5$~TeV of the extra dimensions,
  in $n=2$ to $n=6$ extra dimensions.}\label{fig:dgammademdgraphs}
\end{center}
\end{figure}

Figures~\ref{fig:gammaemdgraphs} and~\ref{fig:dgammademdgraphs}
illustrate the energy profiles of the decay processes for $n=2$ to
$n=6$ extra dimensions.
\begin{figure}
\begin{center}
\includegraphics[width=0.9\columnwidth]{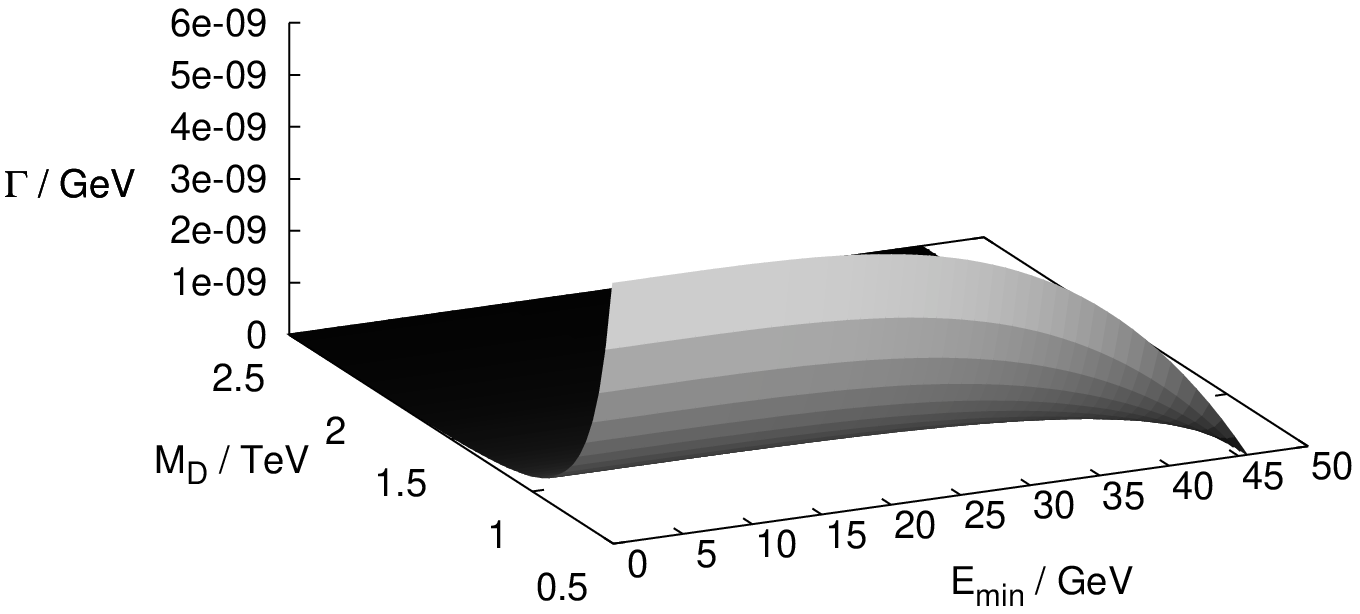}
\\
(a) $n=2$
\end{center}
\end{figure}
\begin{figure}
\begin{center}
\includegraphics[width=0.9\columnwidth]{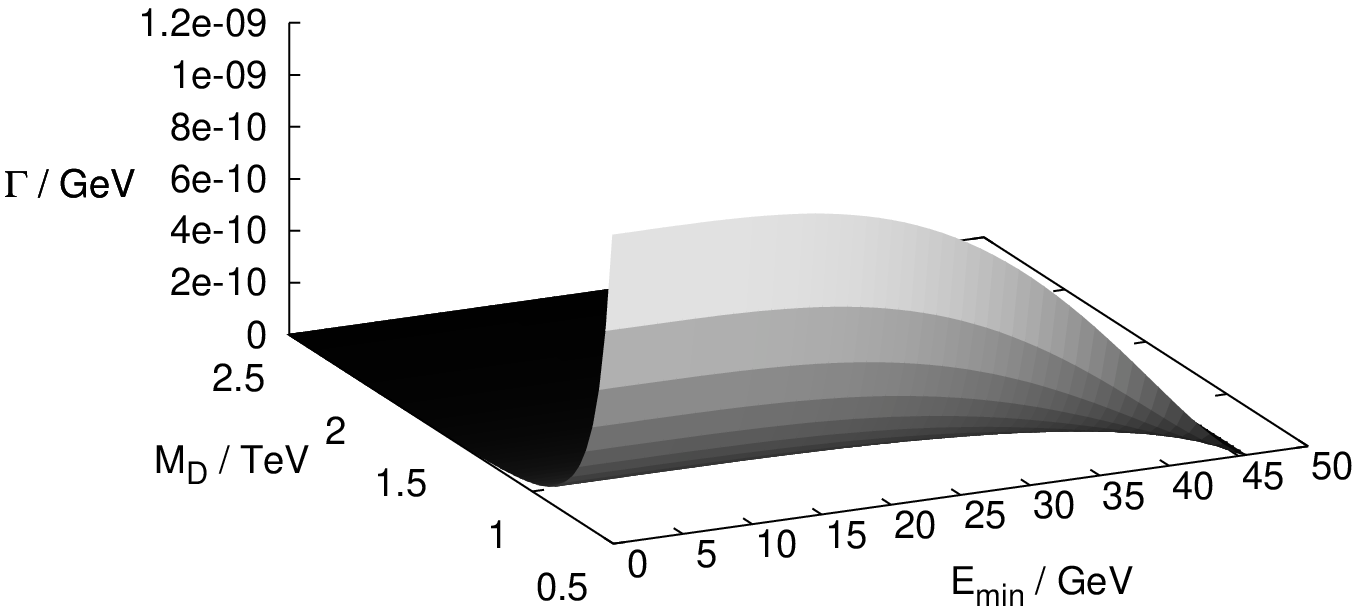}
\\
(b) $n=3$
\end{center}
\end{figure}
\begin{figure}
\begin{center}
\includegraphics[width=0.8\columnwidth]{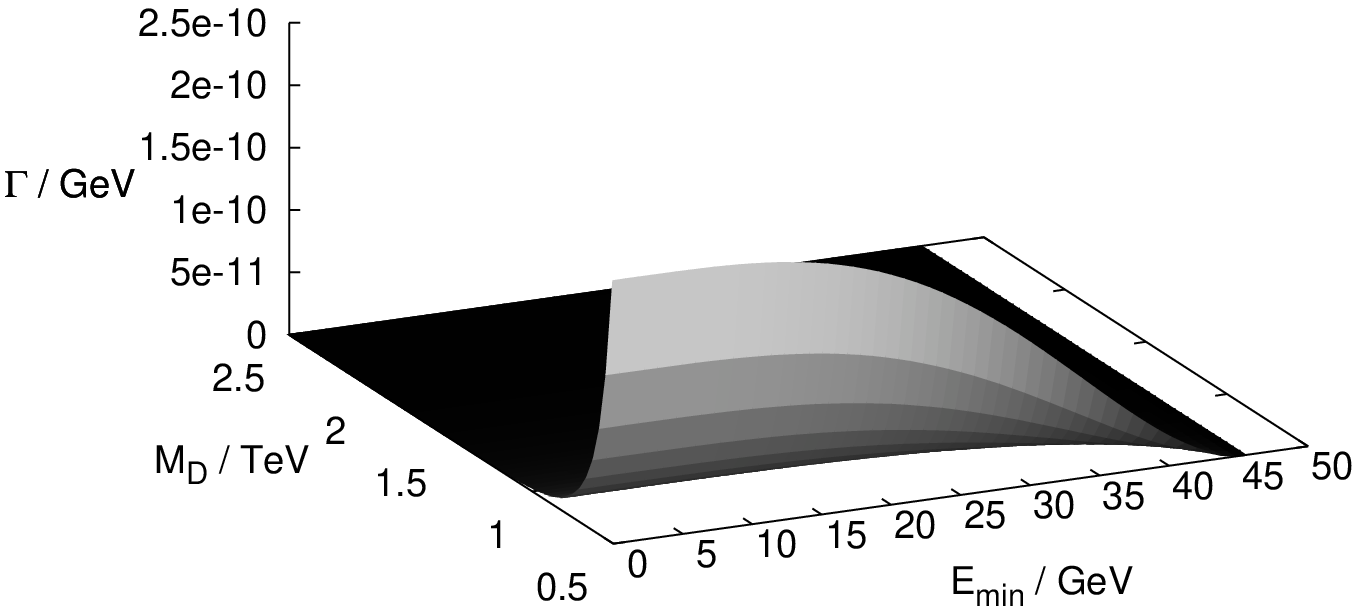}
\\
(c) $n=4$
\end{center}
\end{figure}
\begin{figure}
\begin{center}
\includegraphics[width=0.9\columnwidth]{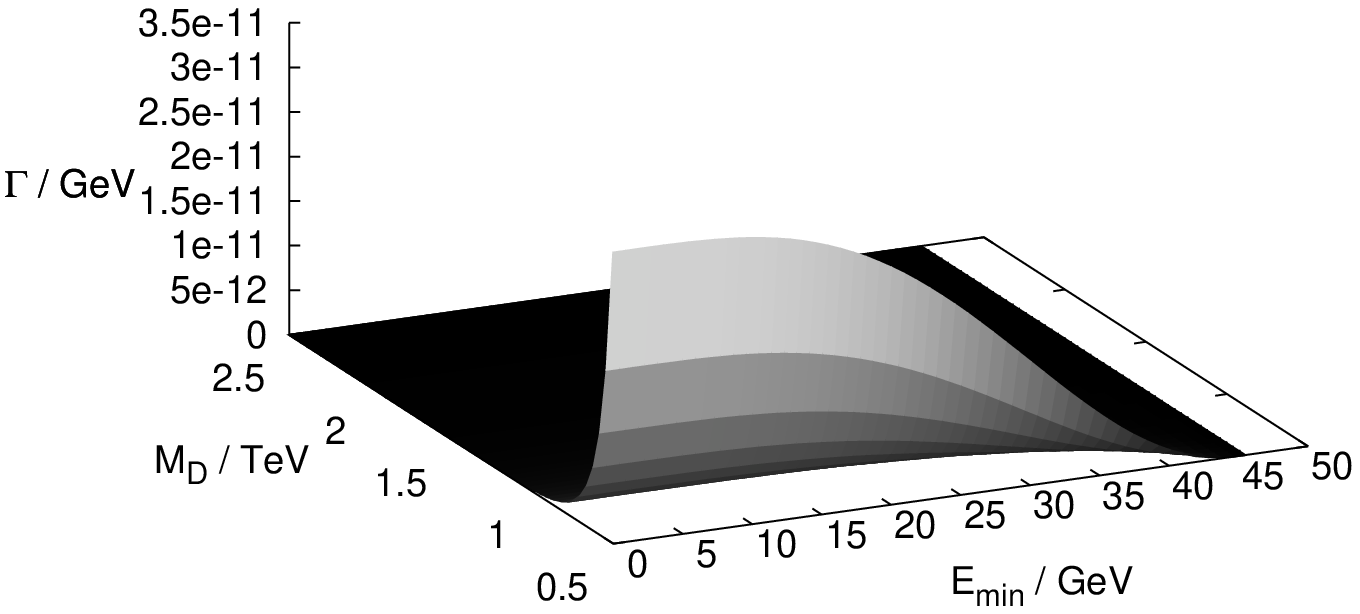}
\\
(d) $n=5$
\end{center}
\end{figure}

It is most obvious from the general profile plots that if such energy
profiles could be obtained from an experiment and there were an
indication of a toroidally-compactified ADD model, it should be
possible to distinguish between $n=2$ and $n>2$ extra dimensions
because of the non-zero energy derivative as $E\to M_Z/2$ in the $n=2$
case.
\begin{figure}
\begin{center}
\includegraphics[width=0.9\columnwidth]{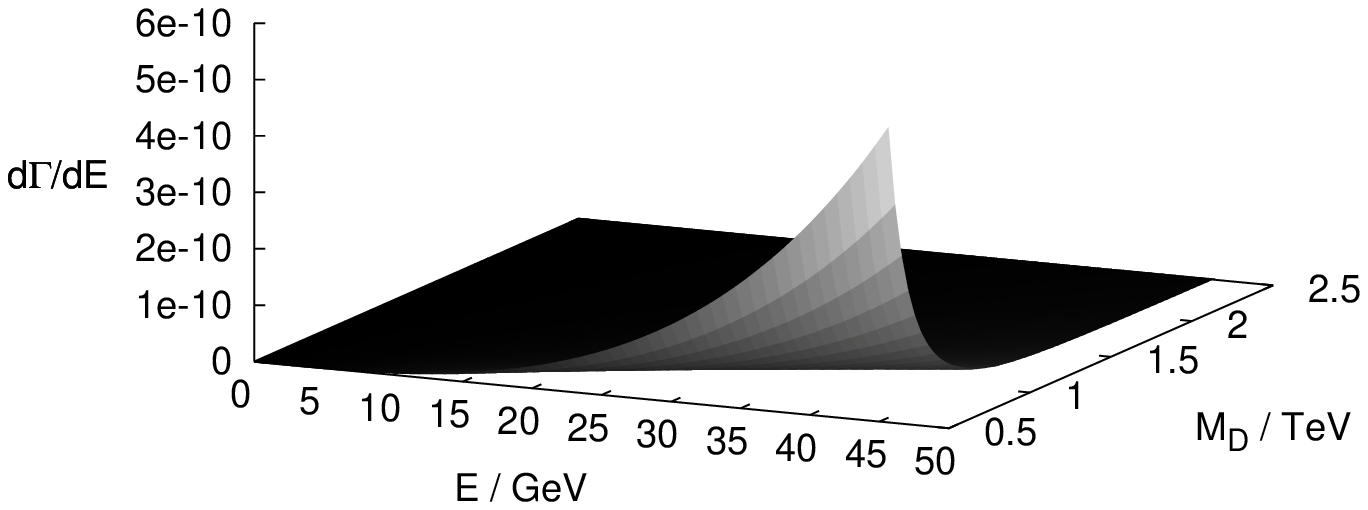}
\\
(a) $n=2$
\end{center}
\end{figure}
\begin{figure}
\begin{center}
\includegraphics[width=0.9\columnwidth]{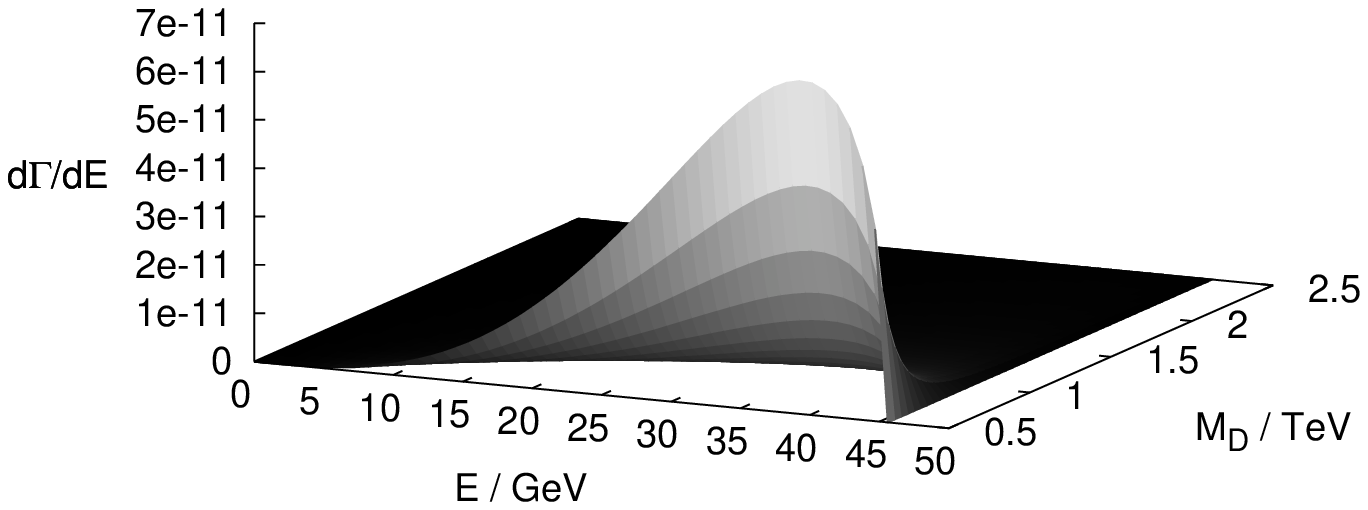}
\\
(b) $n=3$
\end{center}
\end{figure}
\begin{figure}
\begin{center}
\includegraphics[width=0.9\columnwidth]{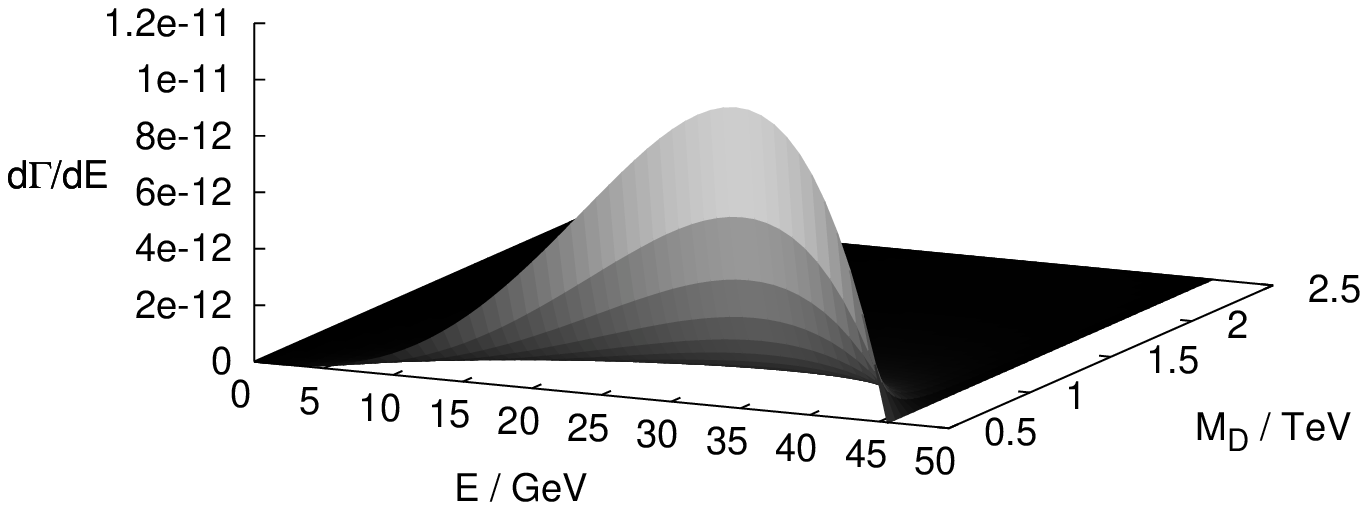}
\\
(c) $n=4$
\end{center}
\end{figure}
\begin{figure}
\begin{center}
\includegraphics[width=0.9\columnwidth]{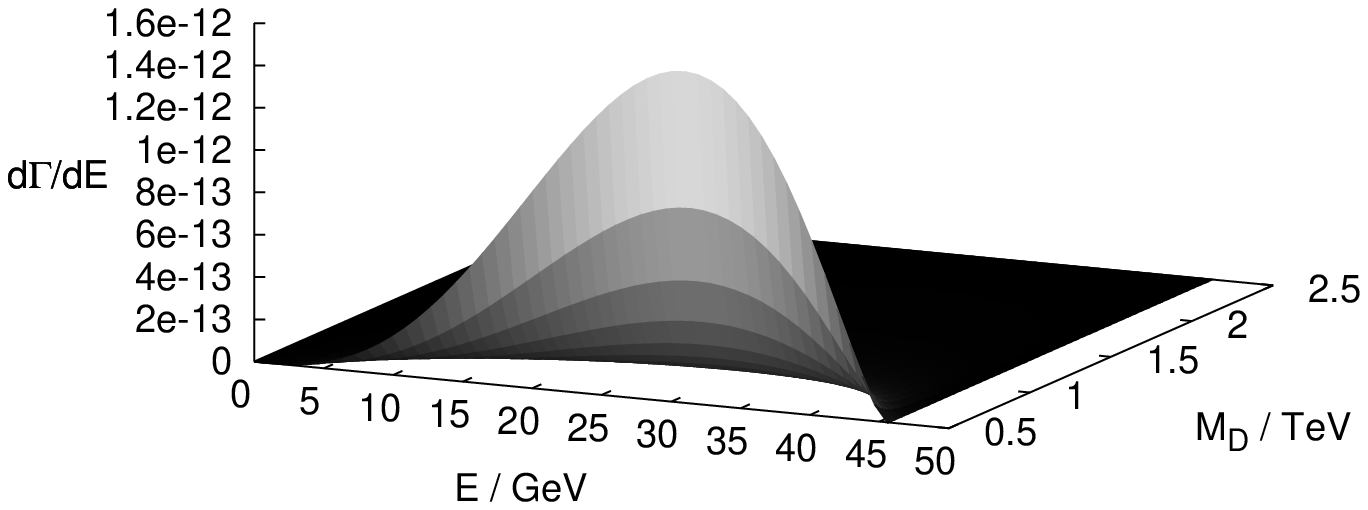}
\\
(d) $n=5$
\end{center}
\end{figure}

\begin{figure}
\begin{center}
\includegraphics[width=1\columnwidth]{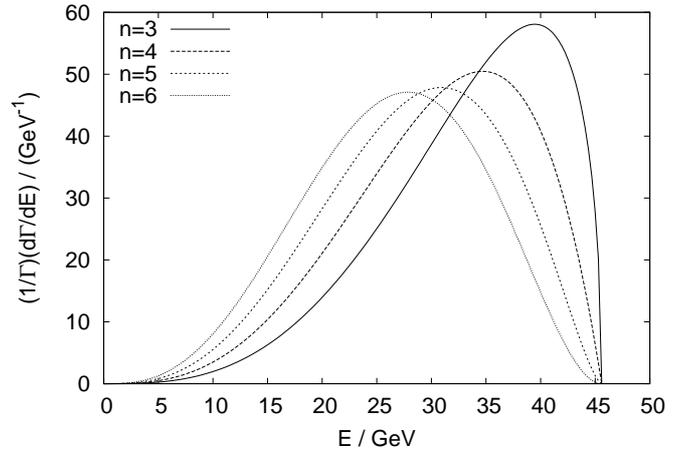}
\caption{Differential energy profiles $d\Gamma / dE$, normalised by
  $1/\Gamma$ to be independent of the extra dimensions scale, for
  $n=3$ to $n=6$ extra dimensions.}\label{fig:normalisedprofilesn3to6}
\end{center}
\end{figure}

It is also possible to see a distinction between the profiles for
other numbers of extra dimensions, if we observe that taking an energy
derivative of equation~\eqref{eq:energydependentwidth} keeps a factor
of $R^n$, so that one can obtain a parameter independent of the scale
of the extra dimensions by considering $(1/\Gamma)\cdot d\Gamma /
dE$. This is illustrated in figure~\ref{fig:normalisedprofilesn3to6}.
A detailed investigation of the number of events required for
distinction between the numbers of extra dimensions is beyond the
scope of this paper.
However, this does show that given a signal of the ADD model, it may be
possible to distinguish between possible numbers of extra dimensions
if it were possible to obtain a sufficiently high number of Z~decays.

Although the decay we are investigating is of a real Z~boson in the
centre-of-mass frame, which has a uniform angular distribution, the
angular distribution does not remain uniform if one takes into account
the production process $e^+e^-\to Z$, and considers the process
$e^+e^- \to Z \to \gamma \mathcal{G}$ at the Z~resonance. In this
case, the decay to a photon and a spin-0 gravi-scalar has a $1+\cos^2
\theta$ angular distribution, and the decay to a photon and a spin-2
KK graviton has a $1+a\cos^2 \theta$ angular distribution, where
$0<a\leq 1$
depends upon the KK mass and attains its maximum value of $1$ in the
limit of a zero KK mass.

We have assumed a uniform distribution in the calculations in this
paper. The L3 data used to derive the novel bounds in this paper has
angular cuts to exclude regions near the beam, meaning that taking the
angular distribution into account would result in our expecting a
smaller number of events (by a factor of about $0.9$) than we should expect assuming a uniform
distribution. This means that the bounds we have derived are stronger
than those that would be derived by taking the angular distribution
into account. (However, even the slightly strong bounds we have
derived are still weaker than bounds from other processes.)

The bounds derived on the branching ratio of $Z\to \gamma \mathcal{G}$ do not
have an angular dependence, but it may be neceassary if using such
bounds to test or exclude this ADD scenario to take into
account the Z~production process in a particular experiment, to
determine whether it is necessary to consider angular distributions.

\section{Conclusions}

We have shown that for a toroidally-compactified ADD model with a common radius for the
extra dimensions, the branching ratio for the process
$Z\to\gamma\mathcal{G}$ is not more than $10^{-11}$ -- sufficiently
small that the process would not be
observed in a Giga-Z collider without a significant luminosity
upgrade. No such decays of on-shell Z~bosons should be expected at the
LHC. This suggests a possible experimental
search strategy in the event that it became necessary to distinguish
ADD from another suspected beyond
Standard Model scenario (namely an investigation of this relatively
``clean'' experimental decay channel -- a significant excess of events would rule
out this ADD model). This branching ratio also
constitutes a test
that ADD would have to fulfil were there some indication of ADD from
one of the other search channels.

\begin{acknowledgement}
We should like to thank K.~Sridhar, David Ward, Bryan Webber, Graham Wilson, and the
members of
the Cambridge Supersymmetry Working Group for helpful
comments made whilst this paper was in preparation. This work has been
partially supported by the United Kingdom's Science and Technology Facilities Council.
\end{acknowledgement}

\end{fmffile}
\end{document}